\documentclass[aps,prl,twocolumn,superscriptaddress,showpacs]{revtex4}

\usepackage{graphicx}% Include figure files
\usepackage{dcolumn} % Align table columns on decimal point
\usepackage{bm}      % bold math
\usepackage{epsfig}

\begin{document}

\title{High pressure sequence of Ba$_3$NiSb$_2$O$_9$ structural phases: new $S = 1$ quantum spin-liquids based on Ni$^{2+}$}

\author{J.~G.~Cheng}
\affiliation{Texas Materials Institute, University of Texas at
Austin, TX 78712, USA}

\author{G.~Li}
\affiliation{National High Magnetic Field Laboratory, Florida State
University, Tallahassee, FL 32306-4005, USA}

\author{L.~Balicas}
\affiliation{National High Magnetic Field Laboratory, Florida State
University, Tallahassee, FL 32306-4005, USA}

\author{J.~S.~Zhou}
\affiliation{Texas Materials Institute, University of Texas at
Austin, TX 78712, USA}

\author{J.~B.~Goodenough}
\affiliation{Texas Materials Institute, University of Texas at
Austin, TX 78712, USA}

\author{Cenke Xu}
\affiliation{Department of Physics, University of California, Santa Barbara, California 93106, USA
}

\author{H.~D.~Zhou}
\email{zhou@magnet.fsu.edu}\affiliation{National High Magnetic
Field Laboratory, Florida State University, Tallahassee, FL
32306-4005, USA}

\date{\today}

\begin{abstract}

By using a high pressure, high temperature (HPHT) technique, the
antiferromagnetically ordered ($T_N$ = 13.5 K) 6H-A phase of
Ba$_3$NiSb$_2$O$_9$ was transformed into two new gapless
quantum spin liquid(QSL) candidates with $S=1$ (Ni$^{2+}$)
moments: the 6H-B phase with a Ni$^{2+}$-triangular lattice and
the 3C-phase with a Ni$^{2+}$-three-dimensional (3D) edge-shared
tetrahedral lattice. Both compounds show no magnetic order down to
0.35 K despite Curie-Weiss temperatures $\theta_{CW}$ of -75.5 K
(6H-B) and -182.5 K (3C), respectively. Below $\sim 25$ K the
magnetic susceptibility of the 6H-B phase saturates to a constant
value $\chi_0 = 0.013 $ emu/mol which is followed below 7 K, by a
linear-temperature dependent magnetic specific heat ($C_M$)
displaying a giant coefficient $\gamma$ = 168 mJ/mol-K$^2$. Both
observations suggest the development of a Fermi-liquid like
ground state characterized by a Wilson ratio of 5.6 in this
insulating material. For the 3C phase, the $C_M \propto T^2$
behavior indicates a unique $S=1$, 3D QSL ground-state.

\end{abstract}

\pacs{75.40.Cx, 75.45.+j, 61.05.C-}% PACS, the Physics and Astronomy

% Classification Scheme.
\maketitle

A quantum spin-liquid (QSL) is a ground-state where strong
quantum-mechanical fluctuations prevent a phase-transition towards
conventional magnetic order and make the spin ensemble to remain
in a liquid-like state \cite{review1, review2}. So far
various gapped spin liquids have been found in dimerized spin
systems and spin ladders \cite{Ruegg1, Ruegg2, spinel, Xu, Ken, Hong, Gaulin, Gardner}.
However, topological and gapless spin liquids are much less well-understood in
dimensions higher than one. Most of the gapless QSL
candidates studied to date are two-dimensional frustrated magnets
composed of either a triangular lattice of $S = 1/2$ dimers, such
as the organic compounds $\kappa$-(BEDT-TTF)$_2$Cu$_2$(CN)$_3$\cite{CuCN1, CuCN2}
(abbreviated as ET) or EtMe$_3$Sb[Pd(dmit)$_2$]$_2$\cite{SbPd1, SbPd2} (abbreviated
as dmit), or of a kagome lattice of Cu$^{2+}$ ($S = 1/2$) ions,
such as the ZnCu$_3$(OH)$_6$Cl$_2$\cite{ZnCu1, ZnCu2}, BaCu$_3$V$_2$O$_8$(OH)$_2$\cite{BaCu},
and the Cu$_3$V$_2$O$_7$(OH)$_2$$\cdot$2H$_2$O\cite{CuV} compounds.

However, whether a gapless QSL can be realized in systems with
larger spins, e.g. $S = 1$, especially in systems with a
three-dimensional (3D) lattice, is still a matter of debate. For
example, the $S = 1$ material NiGa$_2$S$_4$ \cite{NiGaS}with a
triangular lattice develops quadrupolar order
\cite{balentssimon,senthil}, while so far all the 3D gapless
QSL candidates studied to date, such as Na$_3$Ir$_4$O$_8$ with
Ir$^{4+}$ ions \cite{NaIrO}, are either $S = 1/2$ or effective
$S = 1/2$ systems due to strong spin-orbit coupling. Therefore,
the present challenge is to find additional model compounds to
test current theories for gapless QSLs. The key to find a new QSL
candidate is to construct a geometrically frustrated lattice with
specific magnetic ions. A commonly used method to design and
discover new materials is to pursue chemical substitutions,
although the application of high pressures is also an alternative
way to transform crystalline structures and discover new phases
which has not been widely used for synthesizing new frustrated
magnets. Here, we followed the second route to synthesize
frustrated magnets Ba$_3$NiSb$_2$O$_9$ displaying the unique
physical properties shown below.

\begin{figure*}[tbp]
\linespread{1}
\par
\begin{center}
\includegraphics[width=5.0in]{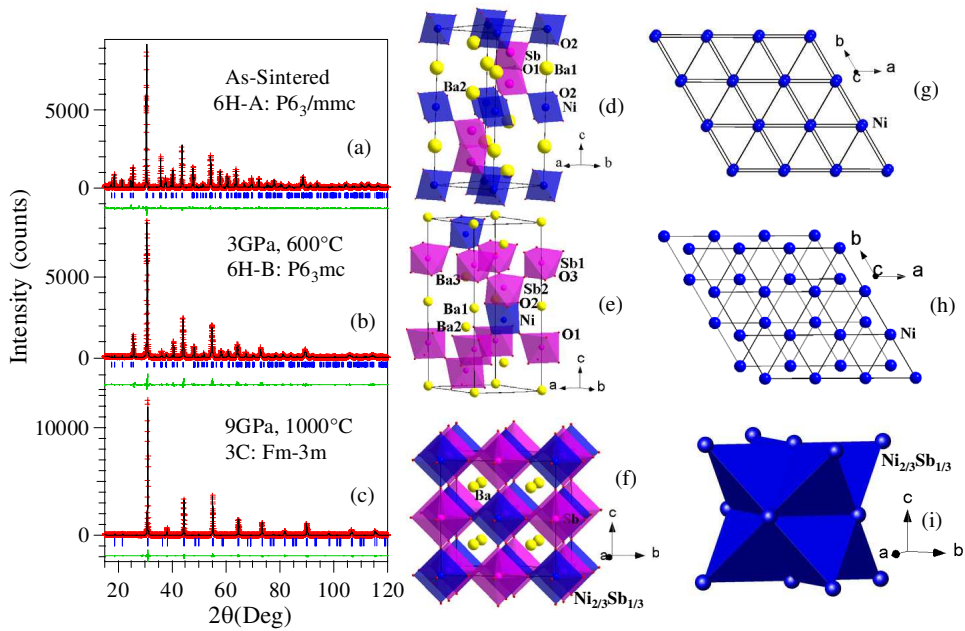}
\end{center}
\par
\caption{(Color online) Powder XRD patterns (crosses) at 295 K for
the Ba$_3$NiSb$_2$O$_9$ polytypes: (a) 6H-A, (b) 6H-B, and (c) 3C.
Solid curves are the best fits obtained from Rietveld refinements using FullProf.
Schematic crystal structures for the Ba$_3$NiSb$_2$O$_9$ polytypes: (d)
6H-A, (e) 6H-B, and (f) 3C, red octahedra represents Sb(M') site and
blue octahedra represents Ni$_{2/3}$Sb$_{1/3}$(M) site.
Magnetic lattices composed of
Ni$^{2+}$ ions for the Ba$_3$NiSb$_2$O$_9$ polytypes: (g) 6H-A,
(h) 6H-B, and (i) 3C.  }
\end{figure*}

The ambient pressure 6H-A phase of Ba$_3$NiSb$_2$O$_9$ was
synthesized through a conventional solid-state reaction. Its x-ray
diffraction (XRD) pattern (recorded at room temperature with Cu
K$_\alpha$ radiation, Fig. 1(a)) shows a single phase having the
hexagonal space group P6$_3$/mmc. The obtained lattice parameters
$a$ = 5.8376(5) {\AA} and $c$ = 14.4013(1) {\AA} agree well with
previously reported values \cite{structure1, structure2}.  The
structure of the 6H-A phase (Fig. 1(d)) consists of dimers of
face-sharing Sb$_2$O$_9$ octahedra linked by their vertices to
single corner-sharing NiO$_{6/2}$ octahedra along the $c$ axis.
The Ni$^{2+}$ ions occupy the 2a Wyckoff site to form a
two-dimensional (2D) triangular lattice in the $ab$ plane (Fig. 1(g)), which is
separated by two non-magnetic Sb layers.

The 6H-B phase of Ba$_3$NiSb$_2$O$_9$ was obtained by treating the
6H-A phase at 600 $^\circ$C under a pressure of 3 GPa for 1 hour
in a Walker-type multianvil module (Rockland Research Co.).
Its XRD pattern (Fig. 1(b)) is different
from that of 6H-A phase and can be satisfactorily indexed as a
distinct hexagonal space group, i.e. the P6$_3$mc with $a$ =
5.7923(2) {\AA} and $c$ = 14.2922(7) {\AA}, respectively. In this
structure (Fig. 1(e)), the dimers of the face-sharing NiSbO$_9$
octahedra (instead of the Sb$_2$O$_9$ octahedra as for the 6H-A
phase) are linked by their vertices to single corner-sharing
SbO$_{6/2}$ octahedra along the $c$ axis. In the well ordered
NiSbO$_9$ octahedra, the Ni$^{2+}$ ions occupy the 2b Wyckoff
sites, which still form a triangular lattice in the $ab$ plane.
For the 6H-A phase, the layers of the Ni triangular lattice are
exactly on top of each other along the $c$-axis. However, for the
6H-B phase, the nearest two layers of the Ni triangular lattice
are displaced with respect to each other in a way that the Ni ion
in one layer is projected towards the center of the triangle
formed by the Ni ions in the adjacent layers along the $c$-axis,
as shown in Fig. 1(h). The instability of the 6H-A phase should
arise from the fact that high pressures tend to reduce the
Sb$^{5+}$-Sb$^{5+}$ distance and therefore partially relieve
strong electrostatic repulsion by exchanging Ni with one of the Sb
atoms. Battle \emph{et al.} reported a similar structure for the
6H-B phase \cite{structure3}, but with no physical
characterization.

With increasing pressure we observed an additional phase
transformation to a cubic perovskite structure. This 3C phase was
obtained under 9 GPa and at a temperature of 1000 $^\circ$C kept
for 30 min. Its XRD pattern (Fig. 1(c)) is best described as a
double-perovskite in a Ba$_2$MM'O$_6$ model with the cubic space
group Fm-3m having a lattice parameter $a$ = 8.1552(2) {\AA}. The
refinement shows a full-ordered arrangement of
Ni$_{2/3}$Sb$_{1/3}$ and Sb atoms at the M and M' sites (Fig.
1(f)), respectively. Therefore the Ni$_{2/3}$Sb$_{1/3}$ sites form
a network of edge-shared tetrahedra, as shown in Fig. 1(i).
Instead of adopting a primitive perovskite structure in which the
Ni$^{2+}$ and Sb$^{5+}$ ions are randomly distributed, the
preferred double-perovskite structure should be attributed to the
large difference in charges between the Ni$^{2+}$ and the
Sb$^{5+}$ ions.

\begin{figure}[tbp]
\linespread{1}
\par
\begin{center}
\includegraphics[width=3.1in]{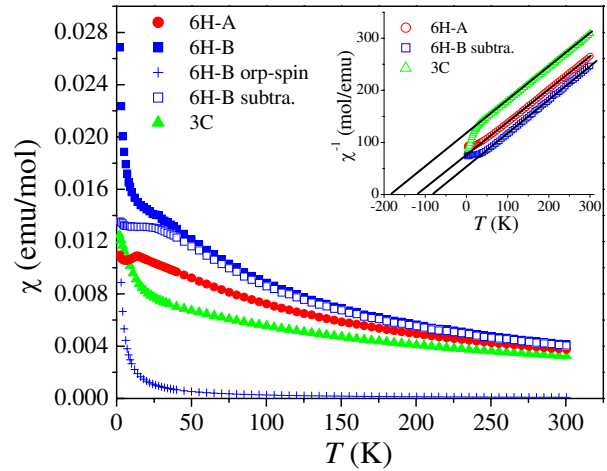}
\end{center}
\par
\caption{(Color online) (a) Temperature dependencies of the DC
magnetic susceptibility ($\chi$) for the Ba$_3$NiSb$_2$O$_9$
polytypes. Inset: Temperature dependencies of 1/$\chi$. The solid
lines on 1/$\chi$ data represent Curie-Weiss fits. For 6H-B phase,
$\chi$ (open squares) is obtained by subtracting 1.7\% Ni$^{2+}$
orphan spin's contribution (crosses) from the as measured data
(solid squares).}
\end{figure}

All three samples are insulators with the room temperature resistance higher than
20 M$\Omega$. The DC magnetic susceptibility ($\chi$($T$), Fig. 2) for all three
compounds was measured under a field \emph{H} = 5000 Oe. For each
compound, one does not observe any difference between the data
measured under zero-field-cooled (ZFC) and that measured under
field-cooled (FC) conditions. The 6H-A sample exhibits a cusp-like
anomaly at the antiferromagnetic ordering temperature $T_N$ = 13.5
K, as previously reported \cite{structure2}. On the other hand,
neither the 6H-B nor the 3C phase show any sign of long range
magnetic order down to 2 K. For the 6H-B phase, we have subtracted
the Curie contribution provided by 1.7 \% Ni$^{2+}$ of orphan
spins from the as measured data. This percentage of Ni$^{2+}$
orphan spins was calculated from fitting the specific heat data\cite{Sup}. After this
subtraction, $\chi$($T$) for the 6H-B phase (open squares in Fig.
2) basically saturates below 25 K with a saturation value $\chi_0$
$\sim$ 0.013 emu/mol. The fittings of the high-temperature region
of $\chi^{-1}$($T$) to the Curie-Weiss law show that all three
compounds have the same value for effective moment,
$\mu_{\text{eff}}$ $\sim$ 3.54 $\mu_B$, as seen from the fact that
all three $\chi^{-1}$($T$) curves are basically parallel to each
other (insert of Fig. 2). This value gives a g-factor of 2.5, which is close to the typical value
for Ni$^{2+}$ ions with spin-orbital coupling\cite{Ni}. The Curie-Weiss temperatures,
$\theta_{\text{CW}}$, obtained for the 6H-A, 6H-B, and 3C phases
are -116.9(4) K, -75.6(6) K, and -182.5(3) K, respectively,
indicating dominant antiferromagnetic interactions for all
compounds.

\begin{figure}[tbp]
\linespread{1}
\par
\begin{center}
\includegraphics[width=3.1in]{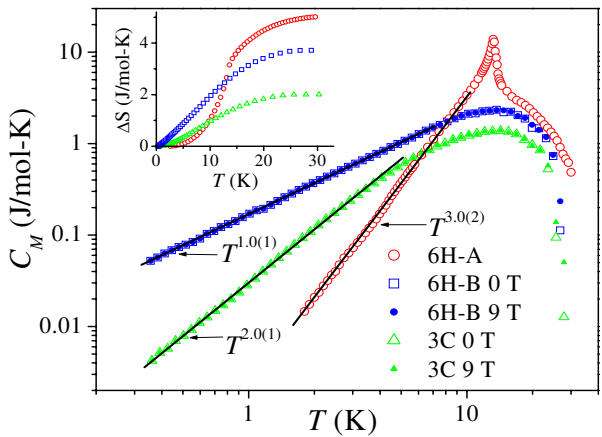}
\end{center}
\par
\caption{(Color online) (a) Temperature dependencies for the
magnetic specific heat (C$_M$) for all three Ba$_3$NiSb$_2$O$_9$
polytypes. Solid lines are the fits as described in the main text.
Inset: variation in magnetic entropy $\Delta S$ below 30 K.}
\end{figure}

The magnetic specific-heat ($C_M$, Fig. 3) for each compound was
obtained by subtracting the heat capacity of the non-magnetic
compound Ba$_3$ZnSb$_2$O$_9$ ordered in the 6H-A, 6H-B, and 3C
phases, respectively, which are used here as lattice standards.
For the 6H-B phase a Schottky anomaly due to 1.7\% of Ni$^{2+}$
orphan spins was also subtracted, see Supplemental
Materials\cite{Sup}. For the 6H-A phase, $C_M$ shows a sharp peak
around $T_N = 13.5$ K. On the other hand, for both the 6H-B and
the 3C phases, $C_M$ which emerges from around 30 K, shows a broad
peak around 13 K with no sign for long-range magnetic-order down
to $T = 0.35$ K. For the 6H-B and the 3C phases, $C_M$ is not at
all affected by the application of a magnetic field as large as
\emph{H} = 9 T.  Below 30 K, the associated change in magnetic
entropy (inset of Fig. 3) is 5.0 J/mol-K, 3.7 J/mol-K, and 2.0
J/mol-K for the 6H-A, 6H-B, and the 3C phase, respectively. These
values correspond respectively, to 55\%, 41\%, and 22\% of $R
\ln(3)$ for a $S = 1$ system, where $R$ is the gas constant. The
remarkable result is that $C_M$ at low temperatures for all three
phases follows a $\gamma T^{\alpha}$ behavior, but with a distinct
value of $\alpha$ for each phase. As shown in Fig. 3, a linear fit
of $C_M$ plotted in a log-log scale yields respectively, $\gamma =
2.0(1)$ mJ/mol-K$^{4}$ and $\alpha = 3.0(2)$ for the 6H-A phase in
the range $1.8 \leq T \leq 10$ K, $\gamma = 168(3)$ mJ/mol-K$^{2}$
with $\alpha = 1.0(1)$ for the 6H-B phase when $0.35 \leq T \leq
7$ K, and $\gamma = 30(2)$ mJ/mol-K$^{3}$ with $\alpha$ = 2.0(1)
for 3C phase within $0.35 \leq T \leq 5$ K.

Both the susceptibility and the specific heat show no evidence for
magnetic ordering down to $T = 0.35$ K for either the 6H-B or the
3C phase, despite moderately strong antiferromagnetic
interactions. The 41\% (6H-B) and the 22\% (3C) change in magnetic
entropy also indicates a high degeneracy of low-energy states at
low temperatures. These behaviors suggest that both the 6H-B and
3C phases are candidates for spin liquid behavior. For the 6H-A
phase, the $C_M \propto T^3$ behavior observed below $T_N$ is
typical for 3D magnons \cite{PrAg}. This indicates that besides
the intra-layer magnetic interactions within the Ni$^{2+}$
triangular lattice, the inter-layer coupling is also relevant for
this phase. As for the 6H-B phase, on the other hand, the relative
shift of the two nearest Ni$^{2+}$ triangular layers leads to a
frustrated inter-layer magnetic coupling, which prevents 3D
long-range magnetic-order. The linear-$T$ dependent $C_M$ of the
6H-B phase is unusual for a magnetic insulator having a 2D
frustrated lattice. Naively, for a 2D lattice one would expect
$C_M$ to display a $T^2$ dependence given by a linearly dispersive
low-energy mode \cite{NiGaS}.

In fact, a series of
recent low temperature studies reveal that $C_M \propto \gamma T$,
with a considerable large value for $\gamma$, is a common feature
among QSL candidates \cite{CuCN1, SbPd4, CuSb}. For example,
ET\cite{CuCN1}, dmit\cite{SbPd4}, and
Ba$_3$CuSb$_2$O$_9$\cite{CuSb}, all composed of a $S = 1/2$
triangular lattice, display $\gamma$ = 12.0 mJ/mol-K$^2$, 19.9
mJ/mol-K$^2$, and 43.4 mJ/mol-K$^2$, respectively.
It has been proposed theoretically that
magnetic excitations or quasiparticles called spinons can lead
to a Fermi surface even in a Mott insulator, which yields a linear term
in the specific heat after the U(1) gauge fluctuation is suppressed due to partial
pairing on the fermi surface\cite{theory1}. The observation of a saturation in
$\chi(T)$ for 6H-B phase enables us
to calculate the Wilson ratio, $R_W$ =
[4$\pi^2$$k_B$$^2$$\chi_0$]/[3($g$$\mu_B$)$^2$$\gamma$]. One
obtains a value of 5.6 by using $\chi_0$ = 0.013 emu/mol and
$\gamma$ = 168 mJ/mol-K$^{2}$. In metals, a Pauli-like
paramagnetic susceptibility and a linear-$T$ dependent heat
capacity, as seen for the 6H-B phase at lower temperatures, which
leads to a concomitant $R_W$ in the order of unity, are
conventional properties of Fermi-liquids. Therefore, we are lead
to conclude that coherent fermionic like excitations or
quasiparticles, which in an insulator can only be magnetic in
nature such as the spinons, are responsible for the low
temperature behavior of the 6H-B phase.

It is known that the ground state of quantum $S = 1$ magnets
depends on the detailed competition between Heisenberg and
biquadratic spin couplings \cite{balentssimon,senthil}. Therefore,
although NiGa$_2$S$_4$ and the 6H-B phase both have similar
triangular lattice structure and spin-1 on each site, their ground
states can be very different. The ground state of NiGa$_2$S$_4$
has quadrupolar order \cite{balentssimon,senthil} with $C_M
\propto T^{2}$ , while in Ba$_3$NiSb$_2$O$_9$ we believe it is the
spinon fermi surface that leads to the constant $\chi$ and
$\gamma$ at zero temperature.

For the 3C phase, despite diluting non-magnetic Sb$^{5+}$ ions
on the Ni sites, the  magnetic interactions between Ni$^{2+}$ ions
are still moderately strong indicated by a $\theta_{\text{CW}}$ = -182.5 K.
The spin liquid like ground-state
then is possibly led by the geometrically frustrated edge-shared
tetrahedra composed of Ni$_{2/3}$Sb$_{1/3}$ sites. The 3C
phase crystallizes in a 3D instead of a 2D lattice, and therefore
the $T^2$ dependence observed for $C_M$ is also unconventional.
Na$_{4}$Ir$_{3}$O$_{8}$, with a 3D $S = 1/2$ (Ir$^{4+}$)
hyperkagome lattice \cite{NaIrO}, also displays a $C_M \propto
T^2$ behavior at low $T$s, which is claimed to be strong evidence
for a QSL ground state and is explained in terms of a spinon
Fermi-surface which is unstable against a spinon pairing state
with line nodes at low energies\cite{theory2}. A similar scenario could be pertinent for the 3C
phase with a face centered cubic structure in which the
nearest-neighbour exchange interaction $J_1$ between the [000] and
[1/2, 1/2, 0] spins on the network of edge-sharing tetrahedra, is
the dominant interaction with a second-neighbour exchange
interaction $J_2$ along the [100] wave-vector being the weaker
one. Former studies on the double-perovskite Ba$_2$MM'O$_6$
already showed that the competition between $J_1$ and $J_2$ can
lead to interesting frustrated magnetic ground states, such as the
valence bond glass state in Ba$_2$YMoO$_6$ \cite{YMo}.

Gapless QSLs are claimed to exist in 3D lattices, or be composed of
spins larger than $S = 1/2$.
For instance, for spin-$S$ systems, the spin model can be
tuned to a SU(2$S$+1) invariant point, where quantum fluctuation
is significantly enhanced, and the semiclassical spin order is
suppressed. Here, we revealed two unique QSL candidates: (i) the 6H-B
phase of Ba$_3$NiSb$_2$O$_9$ having a $S = 1$ moment on a
triangular lattice and displaying $C_M \propto \gamma T$ with a
giant $\gamma$ = 168 mJ/mol-K$^2$, therefore suggesting the
realization of a ground state with a possible spinon fermi
surface but on a $S = 1$ system; and (ii) the 3C phase of
Ba$_3$NiSb$_2$O$_9$ with a edge-shared tetrahedral lattice,
displaying $C_M \propto \gamma T^2$, thus suggesting a rare
example of a 3D-QSL composed of $S = 1$ moments. Most importantly,
we proved that quantum fluctuations are also particularly relevant
for $S=1$ geometrically frustrated systems, which are observed to
follow physical behavior either predicted for or observed in
geometrically frustrated systems composed of $S = 1/2$ moments.

\begin{acknowledgments}
We acknowledge P. A. Lee for discussions, and the critical reading
of this manuscript. This work was supported by NSF (DMR 0904282,
CBET 1048767) and the Robert A Welch foundation (Grant F-1066).
The NHMFL and therefore H.D.Z. is supported by NSF-DMR-0654118 and
the State of Florida. L.B. is also supported by DOE-BES through
award DE-SC0002613.
\end{acknowledgments}

\end{document}